# An Approach of Replicating Multi-Staged Cyber-Attacks and Countermeasures in a Smart Grid Co-Simulation Environment

*Ömer Sen[1*], Dennis van der Velde[1], Sebastian N. Peters[2], Martin Henze[3]*

[1]*Digital Energy, Fraunhofer FIT, Aachen, Germany*
[2]*IAEW, RWTH Aachen University, Aachen, Germany*
[3]*Cyber Analysis & Defense, Fraunhofer FKIE, Wachtberg, Germany*
* oemer.sen@fit.fraunhofer.de

## Abstract

While the digitization of power distribution grids brings many benefits, it also introduces new vulnerabilities for cyber-attacks. To maintain secure operations in the emerging threat landscape, detecting and implementing countermeasures against cyber-attacks are paramount. However, due to the lack of publicly available attack data against Smart Grids (SGs) for countermeasure development, simulation-based data generation approaches offer the potential to provide the needed data foundation. Therefore, our proposed approach provides flexible and scalable replication of multi-staged cyber-attacks in an SG Co-Simulation Environment (COSE). The COSE consists of an energy grid simulator, simulators for Operation Technology (OT) devices, and a network emulator for realistic IT process networks. Focusing on defensive and offensive use cases in COSE, our simulated attacker can perform network scans, find vulnerabilities, exploit them, gain administrative privileges, and execute malicious commands on OT devices. As an exemplary countermeasure, we present a built-in Intrusion Detection System (IDS) that analyzes generated network traffic using anomaly detection with Machine Learning (ML) approaches. In this work, we provide an overview of the SG COSE, present a multi-stage attack model with the potential to disrupt grid operations, and show exemplary performance evaluations of the IDS in specific scenarios.

## 1 Introduction

In view of the ongoing digital transformation of electric power systems to SGs, where complexity and dependence on Information and Communication Technology (ICT) have already increased significantly, new challenges for secure and stable grid operation arise [1, 2]. As a result, more operational processes of power grids are controlled by computers with sensors, monitoring, management services, and high-level automation, i.e., OT, leading to the emergence of (partially) remotely controllable systems with full operational capability without human intervention. Disruptions or large-scale power outages not only cause public chaos, dislocation, and lost production, but can also endanger human lives, as in Venezuela in 2020 [3], and must be protected both physically and digitally [4]. It is known from ICT systems in other domains, that more complexity, more users, and more communication interfaces lead to a larger attack surface for cyber-attacks [5]. Cyber-attacks on critical infrastructure from recent years show the risks for industrial control systems such as Supervisory Control and Data Acquisition (SCADA) systems [4]. The development, validation, and testing of data-driven countermeasures such as ML-based detection methods for sophisticated cyber-attacks depend on the availability and quality of attack data under realistic conditions, generated either synthetically or under representative laboratory conditions, as they are not publicly available [6]. Bridging this gap requires replicating attack scenarios in an isolated, secure, and controllable environment that provides not only valid properties within the energy domain, but also ICT. Thus, to rationally and comprehensibly represent the behavior of a multi-stage attack, the generation of valid attack data based on a structured and plausible approach within a scalable and flexible environment that realistically replicates all necessary layers of an SG application is required [6]. Our aim in this work is to provide an approach to replicate such scenarios to generate the data that can be used to develop, validate, and test domain-specific countermeasures against cyber-attacks. To this end, this paper presents a simulation-driven SG testing environment that aids in the investigation of coordinated cyber-attacks and the development of appropriate remediation and countermeasures. Our contributions are:

(1) We propose our approach to replicate cyber vulnerabilities of OT components and multi-stage cyber-attacks in a COSE.
(2) We show and describe a structured and comprehensive setup for generating cross-domain attack traces.
(3) We present and discuss the demonstrated use case of exemplary deployment and evaluation of countermeasures such as (semi-)supervised ML-based IDS within the SG domain.

## 2 Components & Security in Smart Grids

As a basis for our work, we provide a brief overview of SG core components, their security issues in SG utilities, and highlight possible remediation approaches.

### 2.1 ICT in Power Grids

ICT in power grids enables dynamic control of power generation, consumption, and storage by using advanced control systems to monitor, protect and automatically optimize grid operation [7]. As power fluctuations generally increase due to volatile Distributed Energy Resources (DERs), networking via ICT is becoming an integral part of SGs [8]. The increased use of ICT is causing a paradigm shift in distribution grids, where in the past distribution grid operators had hardly any control and knowledge about distributed power injection into their grid [7]. The conventional control structure between higher-level control units and field- or station-level devices is based on the SCADA, which describes the organization of large spatially distributed ICT systems [9]. Its composition in the energy domain usually consists of a Master Terminal Unit (MTU) with connected Human Machine Interface (HMI) and several Remote Terminal Units (RTUs). The systems are connected via a communication network with routers and switches and separated from the engineering workstations, the company offices and the Internet by a firewall (or even air-gap).

### 2.2 Cyber-Security in Smart Grids

With the growing amount of ICT components in power grids, new security concerns arise due to the relatively long life cycle of IT components in power grids, the homogeneous environment of commercial IT products and systems, the provision of open interfaces, and the increasing interconnection of different actors [10]. This increased interconnectedness can create vulnerabilities, described as flaws or weaknesses in the design, implementation, or operation and management of a system, component, or protocol that can be exploited to violate security policies [11]. In terms of attack propagation, i.e., lateral movement, vulnerabilities can be exploited to control remote systems and networks such as Remote Code Execution (RCE) and Privilege Escalation (PE) [12]. An RCE vulnerability allows user-supplied input to software to be executed on the system by a programming language parser [13]. PE occurs when a user is given more access to resources or functions on a system than they are allowed [13]. Despite a variety of preventive measures such as strict user management, password policies, access control, and network segmentation, intrusion detection capabilities are still needed to meet the high-security requirements in SG. A promising approach to detect such invasive attack propagations is an IDS, which describes a technique for detecting unauthorized access, such as intrusion attempts, to a computer system or network by using either anomaly-based or signature-based approaches, or a mixture of both [14]. In particular, the anomaly-based detection method is based on an approach that classifies observed events within the system being monitored according to the specified or learned system behavior characterization under normal operating conditions. Any deviation from the normal system behavior characterization is classified as an anomaly. For the characterization of normal system operating behavior, an ML algorithm can be used that learns from a dataset of examples and can generalize patterns that exist within the dataset after the learning phase is complete [15]. The following ML approaches are considered in this work for separating, clustering, and deciding network traffic into normal and abnormal traffic [16]:

(1) Random Forest (RF) is a supervised classification method that consists of multiple decision trees to classify data.
(2) K-Nearest-Neighbor (KNN) is a supervised ML algorithm used for classification and regression.
(3) The Local-Outlier-Factor (LOF) algorithm is a neighborhood algorithm that calculates a degree of abnormality or "outlierness" for each point in the data set.
(4) The Isolation Forest (IF) algorithm is used to uncover data points that can be partitioned more quickly than others.

A prominent use case of our approach is the provision of an adequate test environment to replicate multi-stage attack scenarios and provide a foundation to investigate countermeasures such as IDS for their suitability to domain-specific deployment scenarios, which are further described in this paper.

### 2.3 Related Work

Various approaches consider the analysis of coordinated attack scenarios using mathematical modeling or cyber-attack trees [17, 18]. Furthermore, various approaches in COSE are being explored that consider hardware-in-the-loop co-simulation or a synthetic framework capable of simulating attack scenarios in SG applications to generate normal and attack data [19, 20]. In exploring countermeasures, other approaches introduce ML-based IDS solutions based on a random forest algorithm to classify communications data in advanced metering infrastructures based on selected features [21]. In this work, we present an SG test simulation environment for modeling a dynamic, multi-stage attacker based on a tree-like decision logic for determining its attack sequences. Furthermore, by deploying isolated processes in Docker containers in our test environment, we can run a variety of services and vulnerabilities, allowing the modeled attacker to dynamically explore the network to exploit found vulnerabilities. With our focus on network traffic in SG, we can evaluate the suitability of semi-supervised approaches within the domain-specific application and also validate the applicability of the generated data from our test environment.

## 3 Methodology

To support ICT security research in SGs, we propose an approach to simulate coordinated cyber-attacks, defensive countermeasures, and realistic network structures and traffic.

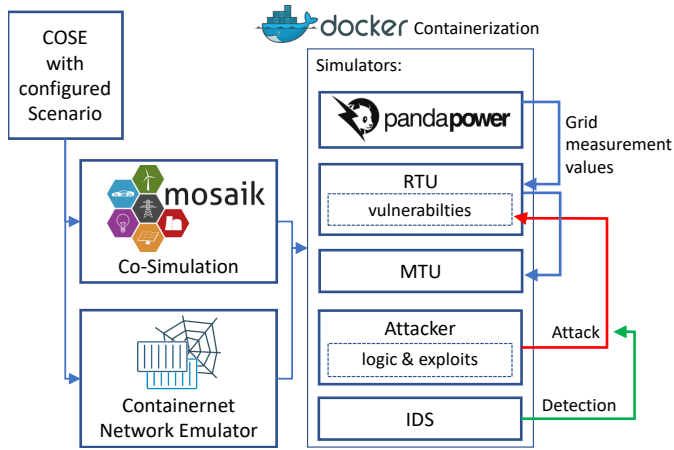

Fig. 1 Overview of the COSE structure, consisting of MOSAIK, Containernet, Pandapower, RTUs, MTU, Attacker, and IDS with their data flow.

### 3.1 Co-Simulation Testbed Setup

Our proposed approach is based on a COSE capable of simulating a Pandapower-based [22] power grid coupled with an IT network that emulates IT and OT devices in a stateful manner managed by a centralized scheduler, i.e., MOSAIK framework [23] (cf. Figure 1). Emulation of IT networks over Containernet [24] enables network transport functionalities between OT devices (e.g., RTU, MTU) via IT (e.g., switches, routers), such as dynamic network structures including protocol conformance, configurable delay, packet loss and bandwidth, and link disconnects. The specification of the simulated infrastructure configuration is based on [2] and all simulators such as OT devices, attackers, and IDSs are based on Docker containers. The focus of this work is on simulating multi-stage cyber-attacks and countermeasures such as IDS in SG. To simulate different attack vectors, we then integrate multiple vulnerabilities into the simulated RTUs with minimal configuration effort, providing exploitation opportunities for multiple stages of an attack (cf. Section 3.2). For conducting observable, multi-stage attacks, we also simulate a decision logic-based Attacker Model (AM) equipped with an arsenal of cyber exploits that provide useful attack data (cf. Section 3.3). The IDS can be instantiated on a device within the IT network emulation environment, capable of analyzing network traffic captured by mirrored SPAN ports (cf. Section 3.4).

### 3.2 Vulnerability Integration

In the context of this work, we address vulnerabilities that allow lateral movement via remote exploitation, such as RCE and PE. The replication of RCE attacks is realized by integrating different types of vulnerabilities covering different protocols such as HTTP/S, Secure Shell (SSH), and Telnet. Many RTUs and industrial computers provide remote access service and setup interface for remote management on certain TCP ports. The concept of these vulnerabilities relies on providing a simple, running service (web interface, SSH server, telnet service) on specific ports (e.g., 80, 22, 23) that are reduced to the ability to execute commands on the computer system (e.g., using known credentials to execute arbitrary commands after logging in). In combination with PE, attack scenarios can be induced during COSE runtime, such as a Denial of Service (DoS) state where simulator traffic is disabled or a compromised simulated device sends spoofed values. After gaining remote access via RCE vulnerabilities, the AM then attempts to escalate privileges on the intruded host (e.g., Linux-based host systems) by exploiting PE vulnerabilities such as set-user-ID (SUID) or sudoers enabled script. Both exemplary privilege escalation vulnerabilities behave similarly to the 'sudo' command on a Linux command line but allow administrative command execution without a password.

### 3.3 Attacker Model (AM) Procedure

Using these vulnerabilities, our AM implements multi-stage attack logic, exploits, and additional software tools whose attack actions and traffic-traces are logged for the purpose of identifying and marking attack data. The attack process itself follows a tree-like structure, starting with a predefined attack goal and ending with the execution of actions to achieve that goal. Thus, the attack setup built for the testbed consists of four stages and focuses on targeting RTU devices:

(1) Find connected systems: Using a network scanner tool, this stage enumerates all reachable systems and their open ports, i.e., all stations (RTUs) found by scanning.
(2) Vulnerability check: Exploit scripts are run for all reachable systems with matching open ports to check if RCE vulnerabilities are accessible behind the ports as described in Section 3.2.
(3) Check privileges and exploit: For all systems compromised in stage 2, user privileges are checked and attempts are made to elevate privileges to become the root user by exploiting PE vulnerabilities as described in Section 3.2.
(4) Affect the target: After gaining the required privileges, actions are chosen to achieve the predefined attack goal (e.g., DoS or data manipulation, e.g., of measurements, to disrupt grid operation).

### 3.4 Intrusion Detection Approach

Within our work, we have chosen the approach of an IDS as a countermeasure to show the capabilities of the COSE in terms of a test environment for the development and testing of detection methods, as well as to demonstrate the usability of the generated data. Various IDS approaches are utilized (cf. Section 2.2) that receive network traffic from a particular RTU switch port through a mirrored port on the central SCADA switch, through which all SCADA traffic in that COSE is routed. Thus, the received traffic sample contains all intended traffic between an MTU and an RTU, as well as potential attack traffic caused by other than intended communications on the network. Captured traffic from the interfaces is collected, decoded, and exported to be analyzed by the selected ML algorithms for anomaly detection (cf. Section 2.2). The utilized high-level data fields are Source, Destination, Protocol, and Length (in bytes). Our generated datasets (scenarios 1-6) contain network traffic under normal and attack conditions with different types of shares (cf. Table 1). The relatively high percentage of attack traffic is due to the generation of large amounts of scan traffic in attack stage 2. Also, scenarios 1-3 are containing a DoS attack, i.e., the RTU stops its service after a successful attack and therefore produces less traffic. Scenarios 4-6 contain a manipulation attack, i.e., the RTU sends manipulated measurement values to the MTU. The data samples are separated into test and training sets and their respective classifications (attack or normal), which are done manually. A data sample can contain both, attack and normal data. E.g., if an attack starts and stops at a certain time of simulation, then the traffic before and after the attack is normal within the same data sample. For supervised ML algorithms, training on classified data is required. The training is performed on a sample containing an attack, while testing is performed with samples containing other attacks. For semi-supervised ML algorithms, unclassified data is sufficient for training, but it must not contain attack traffic to learn normal system behavior.

Table 1 Shares of normal and attack traffic (balance).

| Scenario | 1 | 2 | 3 | 4 | 5 | 6 |
|---|---|---|---|---|---|---|
| Attack [%] | 98.05 | 97.75 | 98.01 | 72.80 | 71.86 | 71.02 |
| Normal [%] | 1.95 | 2.25 | 1.99 | 27.21 | 28.14 | 27.98 |

## 4 Results

In this section, we demonstrate that the COSE is able to replicate different attacks and can be used, for example, to compare different IDS approaches.

*4.1 Attack Replication*

An exemplary multi-stage attack scenario is demonstrated using a simulated medium/low voltage distribution grid equipped with networked assets such as ICT switch, MTU, RTU, DER, etc., (cf. Figure 2). The RTU components are equipped with the vulnerabilities described in Section 3.2. To illustrate further, we also present the automatically executed stages of the attack process within the COSE using terminal executions and a measurement plot of RTU 1 over the simulation steps. In the first stage of the sequence, the AM scans the SCADA network and identifies the connected device, e.g., RTU 1 with its open TCP ports (22, 23, and 80) and configured services (SSH, Telnet, and Nginx webserver). Based on the gathered information, in stage 2 the AM performs RCE via the identified vulnerable web interface on port 80 equipped with a command execution script by specifying the 'whoami' command as a parameter, which issues its executing user 'www-data'. Since the 'www-data' user does not have administrative privileges, the AM extends his privileges in stage 3 through PE by exploiting a found SUID vulnerability. In the final stage, the AM manipulates the measured data about the loading, active and reactive power of the secondary substation transformer and transmits it to the MTU. Thus, it can be suggested that our AM is capable of disrupting grid operations and maintaining this for extended periods of time with sophisticated False Data Injection (FDI) techniques [16].

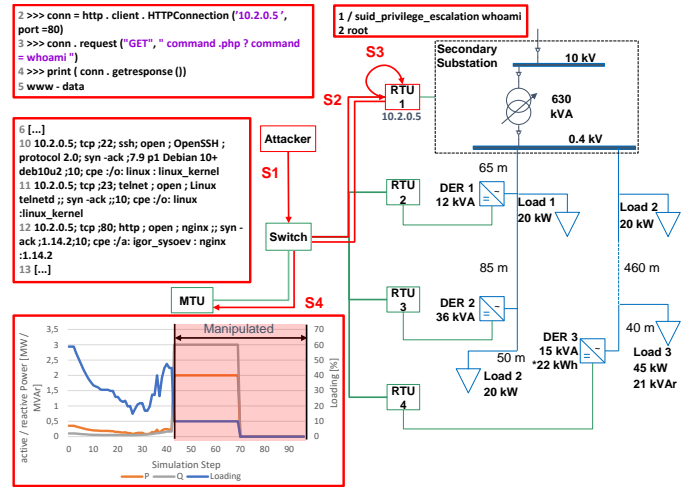

Fig. 2 Illustration of the simulated SG scenario representing a medium/low voltage distribution grid consisting of multiple DERs, resistive/inductive loads, and controlling field devices within a SCADA network. Furthermore, individual attack stages of an exemplary multi-stage attack scenario are visualized, consisting of network scan (S1), RCE (S2), PE (S3), and data manipulation (S4).

*4.2 Countermeasure Evaluation*

Using the deployment and training method of selected ML-based IDS approaches described in Section 3.4, an exemplary evaluation demonstrates the ability of COSE to investigate and compare countermeasures (cf. Figure 3). We evaluate the different ML approaches over the different attack scenarios (cf. Table 1) using the F1 score (harmonic mean of Precision and Recall). For the supervised algorithms, it is observed that KNN and RF achieve nearly equal detection rates for more balanced traffic shares, with RF showing slightly better results within the specified test environment and using only the specified attack sequences. The semi-supervised algorithm LOF produces significantly better detection rates than IF. It can be seen that the attack design and traffic shares have a large impact on the resulting detection quality, with more balanced shares having a more positive effect. However, the absence of a packet is not detected as an anomaly because the algorithms are trained on individual packets and not on packet sequences. To improve detection rates, larger proportions of normal traffic with more complex and diverse attack scenarios are required, for which our COSE provides an advanced foundation to replicate the dynamic nature of SGs with the different data sources, states, and operating conditions exemplified in this work.

## 5 Conclusion

Due to the lack of attack data in critical infrastructure domains, this paper presents an approach to aid the investigation of coordinated cyber-attacks and the development of appropriate remediation and countermeasures such as IDS using a COSE. The COSE has been extended by possibilities for the replication of cyber vulnerabilities and multi-stage cyber-attacks, as well as mutual reachability of devices on the network. It generates protocol-compliant network traffic and takes into account network characteristics such as delays or link disabling. An AM is simulated that can perform network scans, find vulnerabilities, exploit them, gain administrative privileges and execute malicious commands on the OT devices. We demonstrate an example multi-stage attack sequence that has the potential to disrupt grid operations by manipulating measured values. To show the applicability of our approach to improve countermeasure development and validation, we implemented an example of ML-based IDS. It analyzes the generated network traffic using two supervised algorithms (KNN, RF) and two semi-supervised algorithms (LOF and IF). The integrated IDS provides indications of the quality of the generated network traffic data and identifies key challenges for further exploration of this approach. Thus, our COSE-based approach enables the development and investigation of novel countermeasures such as new IDS solutions by providing cross-domain attack data in SG. To provide more diverse and balanced data, future work will include extending vulnerabilities and exploit techniques to different simulated devices, more sophisticated data manipulation strategies such as FDI techniques, and realistic operational logic such as energy management systems.

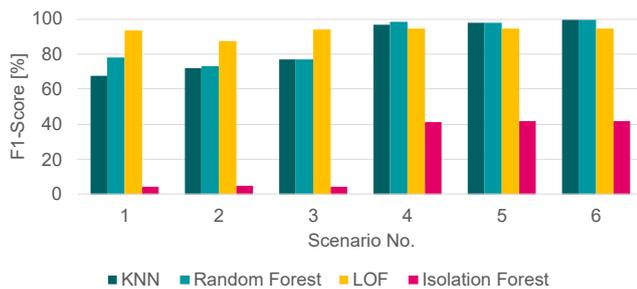

Fig. 3 F1 performance measurement of the ML-based IDS algorithm for 6 different scenarios (cf. Table 1).

*Acknowledgment:* This work has partly been funded by the German Federal Ministry for Economic Affairs and Energy (BMWi) under project funding reference 0350028.